\documentclass[a4paper,twocolumn,11pt,accepted=2019-12-09]{quantumarticle}
\pdfoutput=1
\usepackage[utf8]{inputenc}
\usepackage[english]{babel}
\usepackage[T1]{fontenc}
\usepackage{amsmath}

\usepackage{hyperref}
 
\usepackage{lipsum}

\usepackage{tikz}
\usetikzlibrary{matrix,arrows,decorations.pathmorphing}
\usepackage{tikz-cd}
\usetikzlibrary{arrows}
\usepackage{verbatim}
 \usepackage{authblk}

\usepackage{epsfig}
\usepackage{amsmath,amsfonts,amsthm,amscd,amssymb}
\usepackage{latexsym}
\usepackage{enumerate}

\usepackage{mypack}

\newcommand\change[1]{\textcolor{black}{#1}}

\newtheoremstyle{named}{}{}{\itshape}{}{\bfseries}{.}{.5em}{#1\thmnote{ #3}}
\theoremstyle{named}
\newtheorem*{namedtheorem}{Theorem}

\usepackage[numbers,sort&compress]{natbib}
\begin{document}

\title{Homotopical approach to quantum contextuality}

\author[1,2]{Cihan Okay}
\orcid{0000-0001-8097-5227}
\author[1,2]{Robert Raussendorf}
\orcid{0000-0003-4983-9213}
\affil[1]{Department of Physics and Astronomy, University of British Columbia, Vancouver, BC, Canada}
\affil[2]{Stewart Blusson Quantum Matter Institute, University of British Columbia, Vancouver, BC, Canada}




\maketitle

\begin{abstract}
We consider the phenomenon of quantum mechanical contextuality, and specifically parity-based proofs thereof. Mermin’s square and star are representative examples. Part of the information invoked in such contextuality proofs is the commutativity structure among the pertaining observables. We investigate to which extent this commutativity structure alone determines the viability of a parity-based contextuality proof. We establish a topological criterion for this, generalizing an earlier result by Arkhipov.
\end{abstract}

\section{Introduction} 
Contextuality is a characteristic feature of quantum mechanics \cite{bell64,kochen} that distinguishes quantum from classical physics. It rules out the description of quantum phenomena by so-called non-contextual hidden variable models, in which each observable has a pre-determined value which is merely revealed by the act of measurement. Such models are either internally inconsistent, or fail to reproduce the predictions of quantum mechanics. 

What leads to paradoxes under too stringent assumptions about physical reality becomes a commodity if these assumptions are not made: contextuality is also a resource in quantum computation. For two models of quantum computation---quantum computation with magic states \cite{magic} and measurement-based quantum computation \cite{MBQC}---it has been shown that a quantum speedup can occur only if contextuality is present  \cite{Howard,Nicola,Juan,
AndersBrown,Hoban,RobertContextuality,abramsky2017contextual,
frembs2018contextuality}. 


At its core, contextuality is about the interplay between local consistency and global inconsistency \cite{Abramsky}; and therefore topological methods are particularly suited to discuss it. Presently four approaches to contextuality have been described that employ topology: the topos approach by D{\"o}ring and Isham \cite{DoringIsham,DoringTopos}, the approach based on  \v{C}ech cohomology by Abramsky and coworkers \cite{AbramskyCoho}, the geometric approach by Arkhipov \cite{Arkhipov}, and the approach based on cohomology of chain complexes \cite{Coho}.
For other approaches to contextuality see   \cite{Ana,Spekkens,Cabello,deSilva,roumen2016cohomology} and for refinements of the \v{C}ech cohomology approach see \cite{caru2017cohomology,
abramsky2015contextuality,caru2018towards,Sivert}.



In this paper we unify \change{the  approach of Arkhipov and the approach based on chain complexes}, thereby generalizing one of Arkhipov's results \cite{Arkhipov}.  Both \cite{Arkhipov} and \cite{Coho} are concerned with so-called parity-based contextuality proofs, of which Mermin's square and star are representative examples \cite{Mermin}. Generally, a parity-based contextuality proof is a set of linear constraints on hypothetical value assignments that has no solution.  Although concerned with the same type of contextuality proof, \cite{Arkhipov} and \cite{Coho} use different abstractions and address slightly different questions.

 
Ref.~\cite{Coho} asks: Does a given set ${\cal{O}}$ of observables produce a parity-based contextuality proof? The first step in answering this question is in extracting the essential information of the set ${\cal{O}}$ of observables, in terms of a chain complex ${\cal{C}}$ that captures the commutativity structure within ${\cal{O}}$, and a cocycle $\beta$ that captures the phase information of product relations among commuting elements in ${\cal{O}}$. Then, a parity-based contextuality proof arises in the setting specified by ${\cal{O}}$ if and only if the cocycle $\beta$ is cohomologically non-trivial, $[\beta]\neq0$. 

Arkhipov~\cite{Arkhipov} invokes a different abstraction and addresses a different question. Rather than a chain complex, \cite{Arkhipov} defines an intersection graph $\gG$, with the edges labeling observables and the vertices labeling contexts. Again, $\gG$ captures the commutativity structure among the observables involved. The counterpart to the cocycle $\beta$ of \cite{Coho} is a sign function defined on the vertices of $\gG$. Arkhipov's question is, in a sense, the reverse of the above; namely: Given the intersection graph $\gG$, does there exist a corresponding set ${\cal{O}}$ of observables that produces a parity-based contextuality proof?  This question has a surprisingly simple---and topological---answer. Namely, the intersection graph $\gG$ can give rise to a parity-based contextuality proof if and only if $\gG$ is non-planar.

In \cite{Arkhipov}, rather than a set ${\cal{O}}$ of observables, the intersection graph $\gG$ is the primary object.  \change{The broader significance of this is that a geometrical object encoding  commutativity relations among observables emerges at the base of quantum mechanical contextuality.}
In this paper, we seek generalizations of Arkhipov's results to a greater variety of physical settings \change{and to understand the  structure of the emerging geometric objects}. 
\medskip

We now summarize how the two above-described approaches \cite{Arkhipov} and \cite{Coho} are unified, how one of Arkhipov's results is thereby generalized, and why such a generalization is desirable. First, we observe that the intersection graph $\gG$ employed in \cite{Arkhipov} is defined only for settings in which each observable in ${\cal{O}}$ is part of no more than two contexts. This is a consequence of the simple fact that an edge in a graph has no more than two end points. 

The extension of Arkhipov's results to settings with observables contributing to more than two contexts is, for example, relevant for the discussion of contextuality in measurement-based quantum computation (MBQC).  In the standard MBQC setting \cite{MBQC}, applied to evaluation of Boolean functions on $m$ input bits \cite{RobertContextuality}, there are $2^m$ measurement contexts overall. Each local observable measurable in the process, chosen from a set of 2, appears in  one half of these contexts. Therefore, each such observable appears in more than two measurement contexts whenever $m>2$.
 
Our main result is that one direction of Arkhipov's result can be generalized in the formalism of \cite{Coho} based on a chain complex ${\cal{C}}$, with no bounds on the number of contexts that any given observable can appear in. Specifically, a complex ${\cal{C}}$ can yield a parity-based contextuality proof only if  the topological space $X_\cC$ corresponding to ${\cal{C}}$ is not simply connected, i.e., only if $\pi_1({X}_\cC)\neq 1$. 
\change{For example, in the case of Arkhipov the complex ${\cal{C}}$  can be described by a closed surface $\Sigma_g$ of some genus $g$ together  with a cell structure on it. The intersection graph $\gG$ corresponds to the {\em{dual}} cell structure of $\Sigma_g$. Vanishing of the fundamental group coincides with the planarity of the intersection graph in this case.}
Also, in \cite{Arkhipov} all observables are constrained to have eigenvalues $\pm 1$, whereas the present treatment permits operators with eigenvalues $e^{2\pi i/d}$, for any $d\geq 2$.


\change{
The remainder of this paper is organized as follows. In Section \ref{sec:magic-arrangements} we introduce magic arrangements and explain the connection to the notion of contextuality. Section \ref{sec:topological-realizations} we introduce the basic tools: topological realizations and path operators. Section \ref{sec:simply-connected-realization} contains our main result (Theorem \ref{thm:main}). In Section \ref{sec:simply-connected-realization} we study topological realizations via the fundamental sequence of homotopy groups. Then in Section \ref{RepresentationFundGroup} we move on to define a representation of the fundamental group of a topological realization and use this to refine our main result (Corollary \ref{cor:generalization}). Section \ref{sec:Quantum RealAlgorithmicApp} is on the algorithmic approach \cite{KSpaper,Slo} to parity-based contextuality proofs and how it relates to the topological approach presented in this paper.  
}


\section{Magic arrangements}  \label{sec:magic-arrangements}

Many proofs of contextuality of quantum mechanics employ fixed sets of observables to find obstructions to the existence of non-contextual value assignments. Two key examples are Mermin's square and star \cite{Mermin}. However, the existence of the obstruction does not depend on the precise choice of the observables; for example, conjugating each observable in the set by the same unitary produces an equivalent proof. Rather, it depends only on which subsets of observables commute, and how commuting observables multiply. In Arkhipov's formulation, this information makes up the so-called ``arrangement", and the commutativity part of it the so-called ``intersection graph" of the arrangement. 

The definition of the intersection graph is as follows. Given a fixed set of observables for a contextuality proof, such as Mermin’s square or star, these observables become the edges of the intersection graph, and the sets of commuting observables, i.e., the contexts, become the vertices of the intersection graph. Thus, two edges have the same vertex as end point if and only if the corresponding observables commute. To stick with the above examples, the intersection graphs of Mermin’s square and star (see Fig.~(\ref{fig})) are the bipartite complete graph $K_{3,3}$ and the complete graph $K_5$, respectively. 

Central in \cite{Arkhipov} is the notion of ``magic" arrangements. 
They appear in the context of nonlocal games \cite{nonlocal}, cooperative strategies that demonstrate contextuality as a resource for winning with high probability. 
An arrangement is magic if the edges of its intersection graph can be populated with Hermitian operators with eigenvalues $\pm 1$, but not with numbers $\pm 1$. The significance of this notion is that every magic arrangement gives rise to a proof of contextuality.
Formally, a set of contraints among quantum observables is specified, and a contextuality proof would be a demonstration of the failure to satisfy these constraints by predetermined outcome assignments to the observables. 

The relevant basic notions can be formalized as follows. 
Let $L$ be a finite set, and $\mM$ denote a collection of subsets of $L$. The elements of $\mM$ are called   \textit{contexts}. Together the pair $(L,\mM)$ can be regarded as a hypergraph with vertices $L$ and edges $\mM$. We will   call such a pair an \textit{arrangement}. We will allow operator inverses to appear in the constraints. Therefore we are given a sign function $\epsilon: L\to \set{\pm 1}$. The triple $(L,\mM,\epsilon)$ will be referred to as a \textit{signed arrangement}\footnote{This terminology is different than Arkhipov's.}. We sometimes drop $\epsilon$ from notation and simply write $(L,\mM)$.

Arrangements can be used to define operator constraints. Operator  constraints are determined by a function 
  $\tau:\mM \rightarrow \ZZ_d$  where $d\geq 2$.
 A \textit{quantum realization} of $(L,\mM,\epsilon,\tau)$ is a function $T:L\to U(n)$ for some $n$ such that
\begin{itemize}
\item Each operator satisfies $(T_a)^d=I$.
\item In each context the operators $\set{T_a|\;a\in C}$ pairwise commute.
\item For each context $C\in \mM$ the operators satisfy the constraint
 \begin{equation}\label{eq:constraints}
\prod_{a\in C} T_a^{\epsilon(a)}  = \omega^{\tau(C)}
\end{equation} 
\end{itemize}
where  $\omega=e^{2\pi i/d}$. 
A \textit{classical realization} of $(L,\mM,\epsilon,\tau)$ is a function $c:L \to \ZZ_d$ such that for each context 
 \begin{equation}\label{eq:constraints-cl}
\prod_{a\in C} \omega^{\epsilon(a)c(a) }  =  \omega^{\tau(C)}.
\end{equation} 
We say an arrangement is \textit{(classically) quantum realizable} if $(L,\mM,\epsilon,\tau)$ has a (classical) quantum realization for some $\tau$.
An arrangement is called \textit{magic} if it is quantum realizable but not classically realizable.


When the arrangement satisfies the property that each observable is contained in exacly two contexts one can define the intersection graph.
Can it be judged from the intersection graph of an arrangement alone whether the arrangement is magic? Arkipov answers this question in an  affirmative way, and provides the following characterization.

\begin{namedtheorem}[A] An arrangement is magic if and only if its intersection graph is non-planar.
\end{namedtheorem} 

A shortcoming of this result is that the intersection graph is defined only when each observable appears in no more than two contexts (edges have only two end points). Our generalization removes this restriction: Observables can appear in arbitrarily many contexts.  The geometrical object underlying our construction is a $2$-dimensional cell complex, or more algebraically the associated chain complex. In the special case where Arkhipov’s result applies, i.e., when each observable is only contained in two contexts, then the intersection graph and the chain complex are related by a duality.

\subsection{Operator realization}

Given a signed arrangement $(L,\mM,\epsilon)$ we will introduce the notion of a topological realization. The idea is to encode the operator relations Eq.~(\ref{eq:constraints}) in a topological way.  For this construction we do not require the operators to commute in each context. Therefore we relax the notion of a quantum realization by removing the commutativity requirement.

An \textit{operator realization} for $(L,\mM,\epsilon,\tau)$ is a function $T:L\to U(n)$ such that $(T_a)^d=I$ for all $a\in L$ and satisfies Eq.~(\ref{eq:constraints}) for each context.
\change{In this case for Eq.~(\ref{eq:constraints}) to make sense we order the elements of each context $C$ and write $\set{a_1,a_2,\cdots,a_n}$ for the elements. Then Eq.~(\ref{eq:constraints}) can be written as 
 \begin{equation}\label{eq:constraints-ordered}
\prod_{i=1}^n T_{a_i}^{\epsilon(a_i)}  = \omega^{\tau(C)}.
\end{equation} 
 }

\subsection{Chain complex of an arrangement} We would like to introduce a chain complex before starting our discussion of a topological realization. This algebraic construction associated to a given arrangement   will capture the constraints Eq.~(\ref{eq:constraints}) using a chain complex.

We define a chain complex $\dD_*(\mM)$ whose $2$-chains, $1$-chains, and $0$-chains are given as $\dD_2=\ZZ_d [\mM] $, $\dD_1=\ZZ_d[L]$, and $\dD_0=\ZZ_d$ together with the boundary maps
\begin{equation*}
\begin{split}
 \ZZ_d [\mM] \stackrel{\partial}{\longrightarrow} \ZZ_d[L] \stackrel{0}{\longrightarrow} \ZZ_d & \\\text{ where }\; & \partial[C]=\sum_{i=1}^k \,\epsilon(a_i)[a_i]
\end{split}
\end{equation*}
for each context $C=\set{a_1,a_2,\cdots,a_k}$. The corresponding cochain complex is denoted by $\dD^*(\mM)$. For example, a $2$-cochain is  a function $\mM\to \ZZ_d$, and similarly a $1$-cochain is a function $L\to \ZZ_d$. The coboundary $d:\dD^1\to \dD^2$ is defined as $df(C) = f(\partial [C])$.
We will consider the homology $H_*(\dD(\mM))$ and the cohomology $H^*(\dD(\mM))$ groups.

The relation to the set of constraints given in Eq.~(\ref{eq:constraints}) is as follows: By definition of the cochain complex the function $\tau:\mM \to \ZZ_d$ can be regarded as a $2$-cochain, i.e. an element of $\dD^2(\mM)$. There is a corresponding cohomology class $[\tau]$ in $H^2(\dD(\mM))$. A classical realization of $(L,\mM,\epsilon,\tau)$ is a function $c:L\to \ZZ_d$ that satisfies Eq.~(\ref{eq:constraints-cl}). 
Equivalently, our construction allows us to state this condition as 
$$
dc = \tau
$$ 
i.e. $[\tau]=0$ in the cohomology group.


\section{Topological realizations} \label{sec:topological-realizations}

We can construct a topological space that realizes the chain complex $D_*(\mM)$. The construction can be given as a cell complex consisting of elementary pieces called  $n$-cells. An $n$-cell is an $n$-disk which geometrically corresponds to  vectors of length $\leq 1$ in the Euclidean space $\RR^n$. The construction starts with a collection of $0$-cells i.e. points, and proceeds by gluing the boundary of the cells in a prescribed way --- see Appendix \S \ref{sec:app}. 
 
To realize the chain complex $D_*(\mM)$  we start with a single point, and glue from both end points a $1$-cell for each $a\in L$ to this single vertex. Then for each context $C$ we glue a $2$-cell (disk) whose boundary traverses the loops labeled by $C=\set{a_1,\cdots,a_k}$ with orientation given by the sign function $\epsilon$.
\change{
The ordering of the elements is dictated by the constraints Eq.~(\ref{eq:constraints-ordered}) that is the boundary of the $2$-cell is glued along the labels $a_1,a_2,\cdots,a_n$ respecting the order. 
}
Let us denote the resulting space by $ \Xsing$, emphasizing that it has a single vertex.  Now the chain complex (with $\ZZ_d$ coefficients)   $\cC_*(\Xsing)$ is exactly the chain complex $\dD_*(\mM)$ introduced earlier. For spaces it is customary to denote  the homology and cohomology groups by $H_*(X,\ZZ_d)$ and $H^*(X,\ZZ_d)$, respectively.

Our space $\Xsing$ consists of a single vertex. We will be interested in other spaces that can be obtained by a similar process that is used to construct $\Xsing$ but with possibly using more than one vertex, although the edges and faces are kept the same.
\medskip

\Def{{\rm
\change{
A topological realization of an 
arrangement $(L,\mM,\epsilon)$ is a $2$-dimensional connected cell complex $X$ together with a map $X\to \Xsing$ of cell complexes that induces an isomorphism between the set of $1$-cells and the $2$-cells.
}
}}

\begin{figure}[h!] \center
  \includegraphics[scale=0.7]{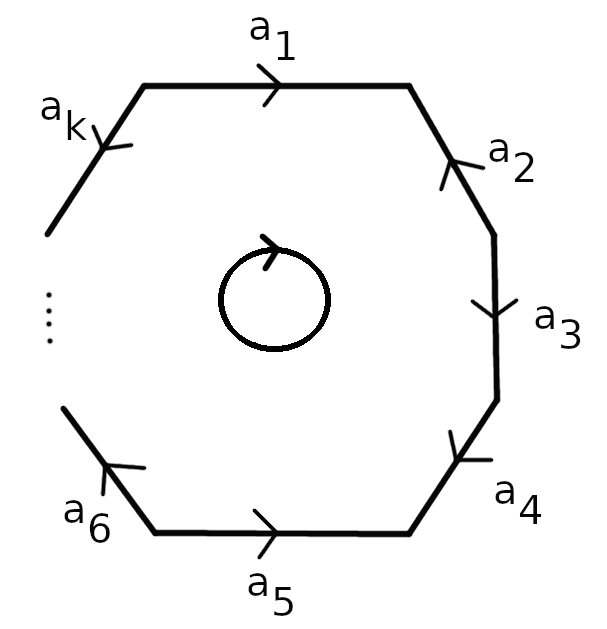}
  \caption{Boundary of a clockwise oriented $2$-cell.}
  \label{fig:twocell}
\end{figure}

 
\change{So} a topological realization of an arrangement is a connected cell complex $X$ which has $1$-cells labeled by $a\in L$ and $2$-cells labeled by $C\in \mM$.   The disk labeled by $C$ has its boundary divided into segments labeled by its elements and oriented by $\epsilon$ (for example as in Fig.~(\ref{fig:twocell})).
In fact, this construction gives a special type of a cell complex, namely, a combinatorial cell complex. \change{This means that  the attaching map of a $2$-cell maps an edge on the boundary homeomorphically to the corresponding edge on the $1$-skeleton.}
 See Appendix \S \ref{sec:app} for a definition of a combinatorial cell complex, and basic properties. 
By construction there is a map $\cC_*(X)\to \dD_*(\mM)$ of chain complexes. Since the $1$-cells and $2$-cells are isomorphic this map induces an isomorphism on second cohomology. As a consequence we can regard the class $[\tau]$ as living in $H^2(X,\mathbb{Z}_d)$.
 
To motivate this definition we relate it to Arkhipov's construction of the intersection graph. In the special type of scenarios that Arkhipov considers one can define an intersection graph $\gG$ from the given data of the hypergraph. Then there is a closed oriented surface $\Sigma_g$ of genus $g$ into which the graph can be embedded. The surface with minimum $g$ determines whether $\gG$ is planar or not. The minimum value of $g$ is zero, that is when $\Sigma_g$ is a sphere, if and only if  the graph is planar. The embedding problem can be thought of possible ways of obtaining a surface from the graph by adding faces (gluing disks). Our framework is dual in the sense that given the embedding $\gG \subset \Sigma_g$ we consider the dual graph $\gG'$ and regard the vertices of $\Sigma_g$ as glued faces to this dual graph. Therefore $ X$ corresponds to the surface $\Sigma_g$ with the cell decomposition determined by $\gG'$.  
\vspace{5mm}

\Ex{{\rm \label{ex:graphs} The intersection graph of the Mermin square and star examples correspond to the graphs $K_{3,3}$ and $K_5$. They can be embedded in a torus as displayed in Fig.~(\ref{fig}). \change{Here the boundary of the square region is identified in a way to obtain a torus.   This can be achieved by identifying parallel edges with the same orientation. The torus has a cell structure   given by triangles or rectangular shapes. The graphs (depicted in blue)
live on the torus in such a way that their vertices match one-to-one with the $2$-cells of the torus. These graphs are non-planar, that is the self-intersection cannot be avoided by moving the edges. We can embed them into a torus, but there is no way to embed them into a plane (or a sphere).}
\begin{figure}[h!]  \center
  \includegraphics[scale=0.53]{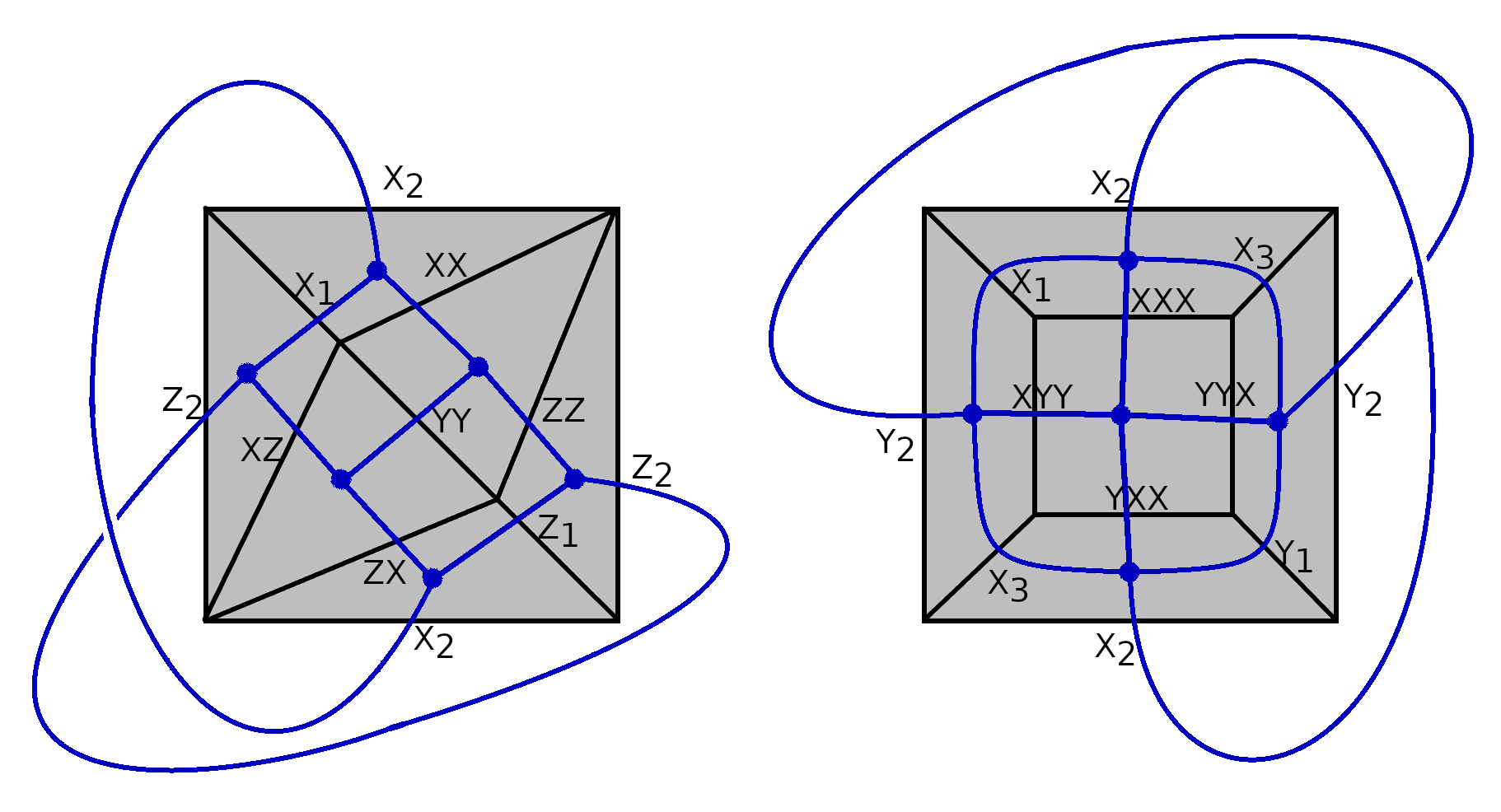}
  \caption{Torus realizing the Mermin square \change{($K_{3,3}$)} and Mermin star \change{($K_{5}$)} arrangements.}
  \label{fig}
\end{figure} 

\change{To illustrate the definition of a topological realization let us consider the Mermin square example.  The triple 
$(L,\mM,\epsilon)$ specifying the arrangement is given by 
$$L=\set{X_1,X_2,XX, Z_1,Z_2,ZZ,XZ,ZX,YY},$$ $\mM$ consists of $6$ contexts given by the $2$-cells. Each context consists of $3$ observables. Our $2$-cells are topologically equivalent to triangles in this case. Observables label the edges of the triangles.  Since each operator squares to identity the sign $\epsilon$ does not have an effect on the operator constraints. However, to be able to specify the topological realization we need to make a choice. We make our choice so that the topological realization gives a torus
 (We will come back to other possible choices, see Fig.~(\ref{fig:mermin-torus-projplane})).   The operator constraints are given by $6$ equations that look like $X_1X_2(XX)=I$, $Z_1Z_2(ZZ)=I$,... Only one of them multiply to $-I$, namely $(XX)(ZZ)(YY)=-I$. Therefore $\tau:\mM\to \ZZ_2$ assigns $0$'s to every  context, except this latter one, which is assigned $1$. Alternatively we can see $\tau$ as a $2$-cochain on the torus, this time interpreted as assigning values in $\ZZ_2$ to each triangle.
 For Mermin star $L$ consists of $10$ observables and $\mM$ consists of $5$ contexts. Each context has $4$ observables. Thus our $2$-cells are rectangular shapes in this case. Similarly $\tau$ assigns values in $\ZZ_2$ to contexts, and the corresponding $2$-cells, determined by the products of the observables in each context.
} 
}}

\subsection{Path operators} Our \change{topological realization} $X$ encodes operator relations by the attaching maps of the $2$-cells. \change{These attaching maps are nice  so that our space is a combinatorial cell complex.} For this purpose \change{in this section we will} assume that $X$ is a combinatorial $2$-complex. See  \S \ref{sec:app} for the definition of combinatorial complexes and combinatorial maps.

Recall that an operator realization is a function $T: L \to U(n)$ and we can think of $L$ as the labels for the edges in a topological realization $X$. 
Fix a vertex $v$ of $X$ for the rest of the discussion. 
For us a path at $v$ is represented by a  combinatorial map $p:S^1\to X$ for some cell structure on $S^1$. In other words, $p$ traverses a sequence of edges $a_1^{\epsilon(a_1)},\cdots,a_k^{\epsilon(a_k)}$ of $X$ where $\epsilon(a_i)=\pm 1$ specifies the orientation. Its image can be just the vertex $v$ as well. We define a path operator
$$
T_p = \prod_{i=1}^k T_{a_i}^{\epsilon(a_i)}.
$$  
 If the image is just $v$ we set $T_p $ to be the identity matrix $I$. 

Given two paths $p,q$ we can define the product path $p\cdot q$ which means traversing $p$ first and then $q$. In the combinatorial language if $q$ is specified by the sequence of edges $b_1^{\epsilon(b_1)},\cdots,b_m^{\epsilon(b_m)}$ then the product would be given by $a_1^{\epsilon(a_1)},\cdots,a_k^{\epsilon(a_k)},b_1^{\epsilon(b_1)},\cdots,b_m^{\epsilon(b_m)}$. By definition of the path operator we have 
$$
T_{p\cdot q}  = T_p T_q.
$$
Another elementary property is that changing orientation of the path amounts to inverting the operator $T_{p^{-1}}=T_p ^{-1}$, again by definition.
\vspace{5mm}

\Ex{{\rm 
Consider two $2$-cells that share a common path as in Fig.~(\ref{fig:cancel}). Let us orient both cells clockwise. Regarding the $2$-cell as $2$-chains the common path $p_2$ gets a plus sign from the left cell and a minus sign from the right cell. The sum of the chains is geometrically represented by the face $F$ obtained by merging the cells. $F$ comes with the clockwise orientation hence it corresponds to the relation that is obtained by multiplying the two constraints
$$
(T_{p_1} T_{p_2})(T_{p_2}^{-1}T_{p_3}^{-1}) = T_{p_1} T_{p_3}^{-1}.
$$ 
\begin{figure}[h!] \center
  \includegraphics[scale=0.9]{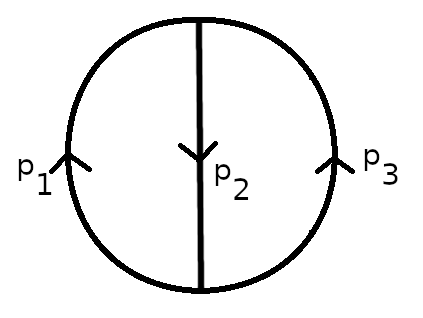}
 \caption{Merging $2$-cells is interpreted as multiplying operator relations. }
  \label{fig:cancel}
\end{figure} 
}}
\vspace{5mm}

\Ex{{\rm
However, there is no one-to-one correspondence between multiplication of operator relations and merging cells (adding chains). For example, consider Fig.~(\ref{fig:exception}), where both $F_1$ and $F_2$ are oriented clockwise. We can cancel the operators in two different ways:
\begin{equation*}
\begin{split}
(T_{p_1} T_b T_{p_2} T_a ) (T_a^{-1}T_{c}^{-1} T_{b}^{-1}T_d^{-1})  & \text{ or }  \\
( T_{p_2} T_a T_{p_1}T_b  ) ( T_{b}^{-1}T_d^{-1}T_a^{-1} T_{c}^{-1})&
\end{split}
\end{equation*}
which possibly gives different results.
The problem is that after canceling $T_a$ (or $T_b$) we cannot cancel the remaining operator.
\begin{figure}[h!] \center
  \includegraphics[scale=0.7]{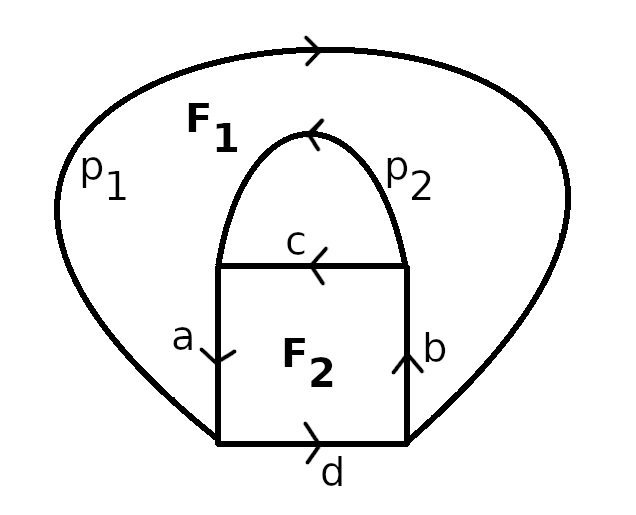}
 \caption{Merging does not correspond to multiplying operator relations.  }
  \label{fig:exception}
\end{figure} 
}}

\section{Simply connected realization} \label{sec:simply-connected-realization}

Given a signed arrangement $(L,\mM)$ we derive consequences about magic arrangement from topological properties of its possible topological realizations. Our main theorem says that if we can realize the signed arrangement on a simply connected (trivial fundamental group) space then the arrangement is non-magic. For topology of $2$-dimensional cell complexes we follow \cite[Chapter II]{TwoDim}.

\subsection{Dragging along a path} We introduce a deformation of a map along a path that will be useful for certain constructions related to homotopy groups. This is a standard trick in algebraic topology.

Let $D^2$ denote the unit disk in $\RR^2$, and $S^1$ denote its boundary. We give a base point $\ast$ to $D^2$ that lies in the boundary, for example we can set $(1,0)$ as  the base point.  
Given a map $g:D^2\to X$ and a path $\beta$ from $w=g(\ast)$ to a fixed vertex $v\in X$ we define a new map $\tilde g_\beta :D^2\to X$ that is obtained by deforming $g$ by dragging the point $w$ backwards along the path $\beta$. We refer to  $\tilde g=\tilde g_\beta$ as the deformation of $g$ along $\beta$. \change{If $\alpha$ is the loop at $w$ given by the image of the boundary under $g$ then $\tilde g_\beta$ maps the boundary of the disk to the loop $\beta \cdot \alpha \cdot \beta^{-1}$ at $v$.
For a more technical definition see \cite[page 341]{Hatcher}.}    

Similarly, we can deform a map $f:S^1\to X$ by dragging along a path and write $\tilde f$ for the resulting map.

\subsection{Fundamental sequence} The key topological tool for studying $2$-dimensional cell complexes is the fundamental sequence. Let $X^1$ denote the $1$-skeleton of $X$, the oriented graph on which we glue the disks. There is an exact sequence of homotopy groups
\begin{equation}\label{eq:fundseq}
0\to \pi_2(X) \to \pi_2(X,X^1)  \stackrel{\partial}{\to} \pi_1(X^1) \to \pi_1(X) \to 1
\end{equation}
see \cite[Chapter II, \S 2.1]{TwoDim}.
We describe the groups in this sequence:
\begin{itemize}
\item $\pi_1(X)$ is the fundamental group of $X$.  It is defined with respect to a vertex $v\in X$, and consists of homotopy classes of maps $f:S^1\to X$ that carries the base point $\ast \in S^1$ to $v$. 

\item Second homotopy group $\pi_2(X)$ is the homotopy classes of maps $h:S^2\to X$ where $h(\ast)=v$.

\item The relative homotopy group $\pi_2(X,X^1)$ consists of homotopy classes of maps $g:D^2\to X$ where $g(\partial D^2) \subset X^1$ and $g(\ast)=v$.
\end{itemize}

Exactness means that at each consecutive map $\cdots \stackrel{\alpha}{\to} A \stackrel{\beta}{\to}\cdots$ in the fundamental sequence, the kernel of $\beta$ is equal to the image of $\alpha$. The most interesting map in the exact sequence is the boundary map $\partial$. It is defined on an element $\Phi:D^2\to X$ by restriction onto the boundary: 
$$
\partial[\Phi] = [\Phi|_{\partial D^2}]
$$
hence as a result one obtains a map $S^1\to X^1$ i.e. an element of the fundamental group of $X^1$.

We start by describing $\pi_1(X^1)$.  
Recall that  $X^1$ is an oriented graph. Let $\tT \subset X^1$ be a maximal tree. $\tT$ consists of all the vertices but it avoids loops. In particular, it is contractible. Select a vertex $v\in \tT$ as a base point. For each remaining edge $e_i$, $1\leq i\leq n$, choose paths $p_i$ and $q_i$ in $\tT$ that connects the two vertices of $e_i$, the source and the target, to the vertex $v$. 
 Consider the loop $a_i = p_i \cdot e_i \cdot q_i$ that consists of traversing $p_i$, $e_i$, $q_i$. Then $\pi_1(X^1,v)$ is the free group generated by the homotopy classes of loops $a_i$:
$$
\pi_1(X^1) \cong F(a_1,\cdots ,a_n).
$$  
\vspace{5mm} 

\Ex{\label{ex:TorusProj} {\rm In Fig.~(\ref{fig:mermin-torus-projplane}) we see two different topological realizations for  Mermin square, \change{see Ex.~(\ref{ex:graphs}) for the explanation of the tuple $(L,\mM,\epsilon,\tau)$.} The difference comes form the orientation of the boundary edges. The resulting spaces are distinct: a torus $X_1=T$ and a projective plane $X_2=\RR P^2$. The maximal trees are given by $\tT_1=\set{e_8,e_9}$ and $\tT_2=\set{e_2,e_8,e_9}$. Note that adding an extra edge to these trees creates a loop. The main difference between the two cases is the number of vertices on the boundary. $X_1$ has a single vertex $v_1$ on the boundary and $X_2$ has two given by $v_1,v_2$. In both cases we need to construct the loops $a_i$ form the edges $e_i$ outside the maximal tree.
\begin{itemize}
\item For $X_1$ the loops at $v_1$ can be chosen as follows: 
\begin{align*}
 a_1&=e_1 \\
 a_2&= e_2\cdot e_9 ^{-1} \\
 a_3 &= e_3\cdot e_8^{-1}\cdot e_9^{-1}\\
 a_4&=e_4 \\
 a_5 &= e_5\cdot e_9^{-1} \\
 a_6 &=e_6\cdot e_8^{-1}\cdot e_9^{-1} \\
 a_7 &= e_9\cdot e_8\cdot e_7.
\end{align*}
Therefore $\pi_1((X_1)^1)$ is the free group $F_7$ on $7$ generators, \change{where $(X_1)^1$ is the $1$-skeleton of $X_1$.}

\item For $X_2$ with base point $v_1$ we can choose: 
\begin{align*}
 a_1&= e_1\cdot e_2\cdot e_9^{-1}\\
 a_3&=e_9\cdot e_2^{-1}\cdot e_3\cdot e_8^{-1}\cdot e_9^{-1} \\
 a_4&=e_9\cdot e_2^{-1} \cdot e_4 \\
 a_5 &=e_9\cdot e_2^{-1}\cdot e_5 \cdot e_9^{-1} \\
 a_6 &= e_9\cdot e_2^{-1}\cdot e_6\cdot e_8^{-1}\cdot e_9^{-1} \\
 a_7&=e_9\cdot e_8\cdot e_7.
\end{align*}
In this case \change{for the $1$-skeleton $(X_2)^1$ the fundamental group} $\pi_1((X_2)^1)$ is the free group $F_6$ on $6$ generators.

\end{itemize}   

}}

Next we move to $\pi_2(X,X^1)$.  First we show that each characteristic map $\Phi_j:D^2\to X$, $1\leq j\leq m$, can be interpreted as an element of the relative homotopy group. Choose a path $\beta_j$ in $\tT$ from the point $\Phi_j(\ast)$ to $v$. Let $\tilde \Phi_j:D^2\to X$ denote the deformation of $\Phi_j$ along $\beta_j$. Then  $\tilde \Phi_j$ satisfies the requirements for a map to be in the relative homotopy group. Thus its homotopy class $[\tilde \Phi_j]$ belongs to $\pi_2(X,X^1)$.
\vspace{5mm}

\Ex{{\rm In Fig.~(\ref{fig:mermin-torus-projplane}) the $2$-cells in both cases are labeled by $C_j$. Let us consider the torus case. We think of $C_1$ as a map $\Phi_1: D^2 \to X$ where the base point $\ast\in D^2$ is mapped to $v_1$, the boundary $\partial D^2$ traverses $e_1\cdot e_2\cdot e_9^{-1}$, and the interior of the disk is mapped homeomorphically to the face $C_1$. We orient each face in the clock-wise direction. In this case we do not need to drag the characteristic map along a path since each face contains the base point $v_1$.
In the $\RR P^2$ case there are faces not containing the base point $v_1$, such as $C_2$ and $C_5$. We can connect them to $v_1$ by dragging along the path $e_9$. So after dragging we think of \change{$C_2$} as a map \change{$\tilde \Phi_2:D^2\to X$} whose boundary traverses $e_9 \cdot e_2^{-1}\cdot e_3\cdot e_8^{-1}\cdot e_9^{-1}$. We do a similar construction for $C_5$. 
}}

\subsection{Action of the fundamental group}

There is an action of $\pi_1(X^1)$ on the relative homotopy group. Let $\gamma$ be a loop in  $X^1$ based at $v$ and $\Phi:D^2\to X$ be a map representing an element of the relative homotopy group. Then the map $\gamma\Phi$ obtained from the action of $\gamma$ on $\Phi$ is given by the deformation $ \tilde\Phi_\gamma$ of $\Phi$ along $\gamma$.  
The boundary map $\partial$ 
respects this action in the sense that
\begin{equation}\label{eq:equivariance}
\partial( \gamma \Phi)  = \gamma \cdot \partial\Phi \cdot \gamma^{-1}.
\end{equation}

Now we can conclude our description of the relative homotopy group. The key technical tool is the approximation result given in Lemma \ref{lem:uptohomotopy} of \S \ref{sec:app}. This result reduces considerations to combinatorial maps. As a result the relative homotopy group $\pi_2(X,X^1)$ is generated by
$$
[\gamma\tilde \Phi_j]\;\;\;\text{ where }\;[\gamma]\in \pi_1(X^1),\;\; 1\leq j\leq m.
$$
Combining this with the fundamental exact sequence gives us a presentation of the fundamental group $\pi_1(X)$ as well as a description of $\pi_2(X)$ in terms of the characteristic maps of the $2$-cells of $X$. 
\vspace{5mm}

\Ex{{\rm As we know that the gluing in Fig.~(\ref{fig:mermin-torus-projplane}) produces a torus and a projective plane we can calculate what the fundamental group is in each case:
$$
\pi_1(T)=\ZZ\times \ZZ\;\text{ and }\; \pi_1(\RR P^2) = \ZZ_2.
$$
Now the fundamental sequence is giving us a presentation for these groups. That is we have
$$
F_7/\im(\partial) = \ZZ\times \ZZ \;\text{ and }\;
F_6/\im(\partial) = \ZZ_2.
$$
This can be verified by considering all the relations imposed by the characteristic maps.
}}

\subsection{Evaluation on a disk} We study the evaluation of the coycle $\tau$ on a disk and relate it to the path operator associated to the boundary. We assume that $(L,\mM,\epsilon,\tau)$, which we used to construct $X$, has an operator realization $T:L\to U(n)$.
 
Recall that the $2$-cells of $X$ encodes the operator relations in Eq.~(\ref{eq:constraints}). We express this using the characteristic map. 
\change{Given a map $g:Y\to X$ we write $g^*\tau$ for the pullback of the cocycle. We will apply this construction to characteristic maps.}
Let $\Phi_j:D^2\to X$ denote the characteristic map of the $2$-cell that corresponds to the context $C_j\in \mM$. We rewrite Eq.~(\ref{eq:constraints}) using the path operator notation:
\begin{equation}\label{eq:const-path}
T_{\Phi_j(\partial D^2)} = \omega^{\Phi_j^*\tau(D^2)}.
\end{equation}
Here $\Phi_j^*\tau$ means that the cocyle $\tau$ is regarded as a $2$-cocycle on $D^2$ via the map $\Phi_j$.  Therefore we can evaluate it on the whole disk to obtain the number $\Phi_j^*\tau(D^2)$.
Another point is that the action of a path $\gamma$ on the characteristic map does not change the  operator
\begin{equation}\label{eq:inv-path}
\begin{aligned}
T_{\gamma \Phi_j(\partial D^2)} =&
T_{\gamma\cdot \Phi_j(\partial D^2)\cdot \gamma^{-1}} \\
  =& T_\gamma T_{\Phi_j(\partial D^2)}   T_{\gamma}^{-1} \\
   =& T_{\Phi_j(\partial D^2)}  
\end{aligned}
\end{equation}
as a consequence of Eq.~(\ref{eq:equivariance}) and Eq.~(\ref{eq:const-path}).

The key observation   is a  generalization of Eq.~(\ref{eq:const-path}). Instead of evaluating $\tau$ on a disk using a characteristic map $\Phi_j$ we consider the general case of a cellular map $g:D^2\to X$. We can ask about the result of evaluating the cochain $g^*\tau$ on the whole disk. The result turns out to be exactly the path operator associated to the loop $g(\partial D^2)$.
\vspace{5mm}

\Lem{\label{lem:key}
For a combinatorial map $g:D^2\to X$ we have  
$$
T_{g(\partial D^2)} = \omega^{g^*\tau(D^2)}.
$$
}    

\Proof{
By definition of a combinatorial map, $g$ satisfies the two properties
\begin{itemize}
\item on some $2$-cells $ g$ acts as a (signed) characteristic map $\Phi_{j_l}^\pm $, $1\leq l\leq r$,
\item the remaining $2$-cells are product cells and they are collapsed to a $1$-cell or a $0$-cell.
\end{itemize}
 A product cell (define more precisely in \S \ref{sec:app}) is a rectangular region with horizontal boundary components $h_1,h_2$  and vertical boundary components $v_1,v_2$. Parallel boundary components are oriented in the same direction. Without loss of generality we can assume $g$ collapses the vertical direction of this cell. Therefore $T_{v_1}$ and $T_{v_2}$ are both equal to the identity operator. On the other hand $T_{h_1}=T_{h_2}$. Therefore the operator assigned to the boundary of a product cell is given by the identity operator. We can remove the product cells one by one to obtain \change{
$$
T_{g(\partial D^2)}  = \prod_{l=1}^r T_{ \Phi^{\pm 1}_{j_l}(\partial D^2)}=\prod_{l=1}^r T_{ \Phi_{j_l}(\partial D^2)}^{\pm 1}.
$$   }
 Note that $D^2$ is a compact space, and there are only finitely many cells. Therefore the elimination process eventually terminates.
Using Eq.~(\ref{eq:const-path}) for each $j_l$ gives us the evaluation of $g^*\tau$ on the whole disk since the product cells do not contribute to the evaluation.
 
}

\subsection{Non-magic arrangements} A connected space whose fundamental group is trivial is called a simply connected space \cite{Hatcher}.   We will consider a simply connected topological realization $X$.  Since $X$ is a $2$-dimensional cell complex Hurewicz theorem \cite[Corollary 4.33]{Hatcher} implies that $X$ is homotopy equivalent to a wedge sum of $2$-spheres. Wedge sum means that the spheres are glued as a single point. For a pair of spheres the resulting space is denoted by $S^2\vee S^2$. Therefore we need to understand how $\tau$ evaluates on spheres. So we consider a map $h:S^2\to X$. The cochain $h^*\tau$ defined on the sphere only depends on the homotopy class of $h$. That is if $h'$ is another such map homotopic to $h$ then $h^*\tau=h'^*\tau$. This is the homotopy invariance of cohomology. In the following we will approximate an arbitrary map by a combinatorial map to be able to apply Lemma \ref{lem:key}. \change{Roughly, the idea is to think of a sphere as composed of two disks glued along their boundary and use Lemma \ref{lem:key} for each disk to conclude that evaluation of $\tau$ on each piece is the same as the product of the operators on the boundary. Since the boundary of the disks are oriented oppositely  these products cancel out exactly.} 
\vspace{5mm}
 
\Lem{\label{lem:sphere} For any map $h:S^2\to X$, where $X$ is an arbitrary topological realization, we have $h^*\tau(S^2)=0$.}
 
\Proof{The fundamental sequence Eq.~(\ref{eq:fundseq}) implies that $[h]\in \pi_2(X)$ can be written as an element $[\Phi]$ in the relative homotopy group $\pi_1(X,X^1)$ such that  $\partial[\Phi]=1$ in $\pi_1(X^1)$. Moreover by Lemma \ref{lem:uptohomotopy} we can first deform $\partial\Phi$ to a combinatorial loop, and then we can deform the resulting map to a combinatorial map on the whole disk. Thus $h$ is homotopic to a combinatorial map $\Phi:D^2\to X$ such that $\partial \Phi$ is a contractible loop in the $1$-skeleton $X^1$. 
Since homotopic maps induce the same map in cohomology we have $h^*\tau = \Phi^*\tau$ and applying Lemma \ref{lem:key} to $\Phi$ we obtain
$$
\omega^{h^*\tau(S^2)}= \omega^{\Phi^*\tau(D^2)}= T_{\partial\Phi(D^2)}. 
$$
It remains to calculate $T_{\partial\Phi(D^2)}$.  As a consequence of contractability of $\partial\Phi$ there exists a map $\bar \Phi: D^2\to X^1\subset X$ that extends the combinatorial map $\partial \Phi$. Up to homotopy we can assume $\bar \Phi$ is also  combinatorial (Lemma \ref{lem:uptohomotopy}). The cell structure on $D^2$  consists of only product cells since the image of $\bar \Phi$ is a path. Now we can apply Lemma \ref{lem:key} to $\bar \Phi$:
$$
T_{\partial\Phi(D^2)} =T_{\partial\bar\Phi(D^2)}= \omega^{\bar\Phi^*\tau(D^2)} = I
$$
since product cells do not   contribute to the evaluation of the cocycle.
}
\vspace{5mm} 

Now we come to our generalization of Arkhipov's result on restricted arrangements, i.e. each  observable belongs exactly to two contexts, that    the  arrangement is  magic only if the intersection graph is non-planar.  
 \vspace{5mm}

\Thm{\label{thm:main}If an arrangement $(L,\mM)$ is topologically realizable by a simply connected space $X$  then it is non-magic.
}
\Proof{ Non-magic means that either there is no quantum realization or if there is one then we need to show that it has a classical realization. Suppose that there is a quantum realization for some $\tau$. As we observed above $X$ can be continuously deformed into  a wedge of spheres $  S^2\vee \cdots \vee S^2$. This decomposes the cohomology as 
$$
H^2(X,\ZZ_d) \cong H^2(S^2,\ZZ_d) \oplus \cdots \oplus  H^2(S^2,\ZZ_d) .
$$  
By Lemma \ref{lem:sphere} the cocycle $\tau$ vanishes on each sphere. This implies that $[\tau]=0$, or equivalently, there exists a function $c:L\to\ZZ_d$ such that $dc=\tau$. This function gives the desired classical realization.      
}
\vspace{5mm}
  
\Rem{\rm{In fact Theorem \ref{thm:main} is more general. Stated for operator realizations it says that if $(L,\mM,\epsilon,\tau)$ can be topologically realized by a simply connected space and it has an operator realization then it can also be classically realized.
}  }

\section{\change{Representation of the fundamental group}}  \label{RepresentationFundGroup} 
 
In this section we reinterpret Lemma \ref{lem:key} and produce a projective representation of $\pi_1(X)$. 
This refined approach will allow us to improve Theorem \ref{thm:main}.

\begin{widetext}
\begin{center}
\begin{figure}[h!] \center
  \includegraphics[scale=0.8]{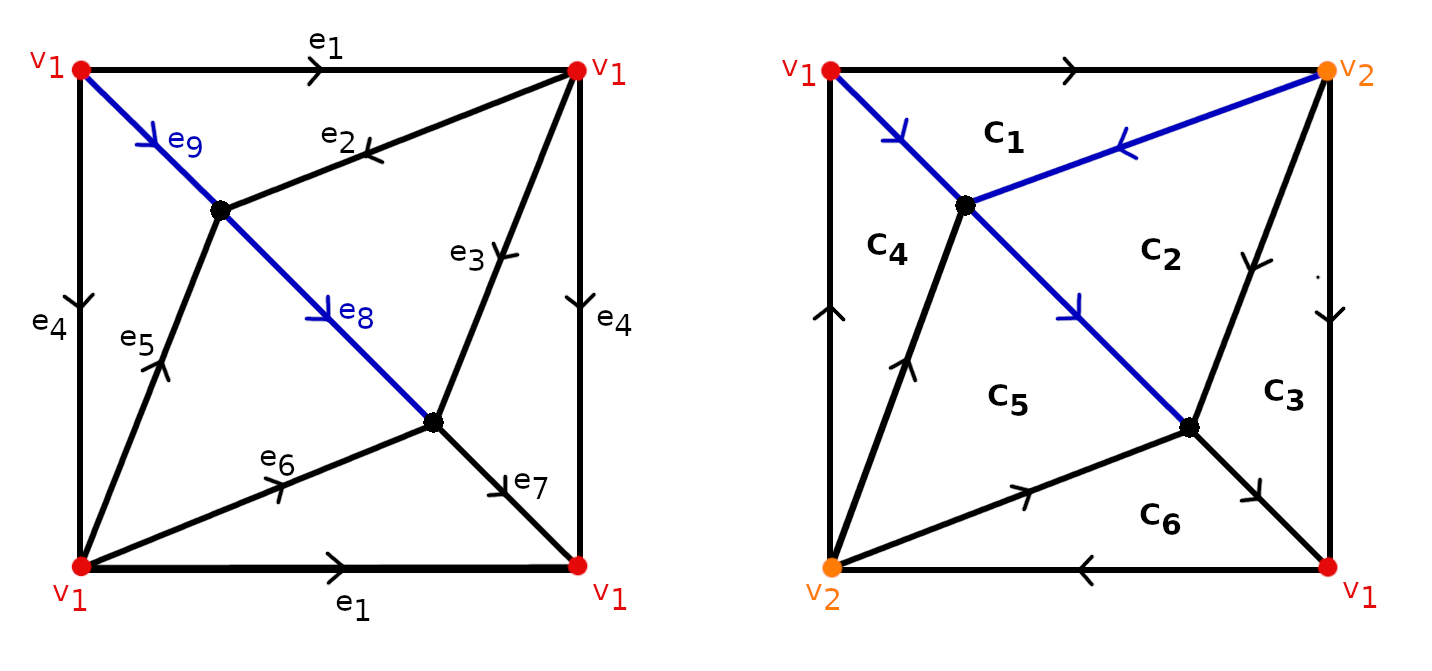}
  \caption{Two different ways of orienting the edges of the surface realizing the Mermin square. Right one is the projective plane $\RR P^2$ and the left one is the torus. Blue edges correspond to a possible choice for a maximal tree $\tT$. The edges are labeled by $e_i$, vertices by $v_k$, and contexts by $C_j$. }
  \label{fig:mermin-torus-projplane}
\end{figure}  
\end{center}
\end{widetext}

\subsection{Definition of the representation} 

\begin{figure}[h!] \center
  \includegraphics[scale=1]{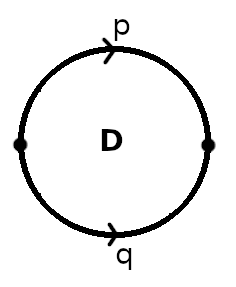}
  \caption{The boundary of the disk is mapped to the two paths that are homotopic.
}
  \label{fig:homotopy-disc}
\end{figure}

Let  $T:L\to U(n)$ be an operator realization for the data $(L,\mM,\epsilon,\tau)$. Given a topological realization $X$ of the signed arrangement $(L,\mM)$ we can consider the assignment
$$
p \mapsto T_p
$$ 
from a loop $S^1\to X$ (combinatorial map) at $v$ to the unitary matrices. Lemma \ref{lem:key} says that if $p$ is homotopic to $q$, written as $p\sim q$, then the disk $H:D^2\to X$ which realizes this homotopy (see Fig.\ref{fig:homotopy-disc}) gives us
$$
\omega^{H^*\tau(D^2)}= T_{H^*(\partial D_2)} = T_{p} T_q^{-1}.
$$
Here the homotopy  $p\sim q$ is thought of as a map from the disk that sends the upper half of the boundary to $p$ and the lower half  to $q$. As a result we obtain that $T_p$ and $T_q$ differ by a scalar, written as $T_p\sim_\omega T_q$.
So modulo the scalar subgroup $\Span{\omega}$ we obtain a well-defined homomorphism 
$$
T: \pi_1(X) \to U(n)/\Span{\omega},\;\;\;\; [p]\mapsto [T_p]
$$
where $[T_p]$ denotes the equivalence class of the operator under $\sim_\omega$.

\subsection{Linearization} An equivalent way of thinking about this representation is to consider   the fundamental sequence (see Eq.~\ref{eq:fundseq}) of $X$:
\begin{equation} \label{diag:representation}
\begin{tikzcd} 
\pi_2(X,X^1) \arrow{r}{\partial} \arrow{d} & \pi_1(X^1) \arrow{d}{T^0}\arrow[r,twoheadrightarrow ,"\rho"] & \pi_1(X) \arrow{d}{T}  \\
\Span{\omega} \arrow[r,hook] & U(n) \arrow[r,twoheadrightarrow ] & U(n)/\Span{\omega}   
\end{tikzcd}
\end{equation}
This makes the construction more concrete since $T^0$ can be chosen more specifically using the isomorphism $\pi_1(X^1)\cong F(a_1,\cdots,a_n)$. We set 
$$
T^0(a_i) = T_{a_i}
$$ 
and extend it to a homomorphism on the free group.  

Given a function  $\alpha:\set{a_1,\cdots,a_k}\to \ZZ_d$ let us define a new homomorphism $T^\alpha: \pi_1(X^1)\to U(n)$ by setting $T^\alpha(a_i)=\omega^{\alpha(a_i)} T^0(a_i)$. \change{Let $\im\, \partial$ denote the image of the homomorphism $\partial$.}
\vspace{5mm}

\begin{lem}\label{lem:linearization}
The projective representation $T:\pi_1(X) \to U(n)/\Span{\omega}$ can be linearized i.e. it lifts to a linear representation $\tilde T$  making the diagram commute
$$
\begin{tikzcd}
& U(n) \arrow{d} \\
\pi_1(X) \arrow[ru,dashed,"\tilde T"]\arrow{r}{T} & U(n)/\Span{\omega}
\end{tikzcd}
$$
if and only if there exists a function $\alpha$ such that $T^{\alpha}(\change{\im\, \partial})=I$.
\end{lem}
\Proof{
Assume that the lift $\tilde T$ exists. Let us write $\bar a_i$ for the image of a generator under the quotient map $\rho:\pi_1(X^1)\to \pi_1(X)$.
Then $\tilde T(\bar a_i)$ has to be $\omega^{\alpha(a_i)} T^0(a_i)$ for some $\alpha(a_i)\in \ZZ_d$ by commutativity of the diagram in Eq.~(\ref{diag:representation}). We can set $T^{\alpha}_{a_i}=\omega^{\alpha(a_i)}T_{a_i}$. Then $T^{\alpha} = \tilde T\circ \rho$ and we obtain that $T^{\alpha}(\im\, \partial) = \tilde T \circ \rho (\im\, \partial) = I$ by exactness i.e.  $\rho \,\partial$ is the identity map.

Conversely, if there exists $\alpha$ such that $T^\alpha$ maps the image of $\partial $ to $I$ that means  $T^{\alpha}$ descends to a homomorphism  $\tilde T:\pi_1(X)\to U(n)$ which turns out to be a lift of $T$. Here we use the presentation of $\pi_1(X)$ given by the upper row of Eq.~(\ref{diag:representation}). 
}

\subsection{Classical realization as linearization} We describe the link between \change{the projective representation of the fundamental group} and the classical realization problem. This result will be used to generalize Theorem \ref{thm:main}.
\vspace{5mm}

\Lem{\label{lem:pathsinX1} Suppose that $p,q$ are two loops in $X^1$ such that $[p]=[q]$ in $\pi_{\change{1}}(X^1)$. Regarding $p$ and $q$ as paths in $X$ we have
$
T_p=T_q.
$}
\Proof{The idea is similar to the proof of Lemma \ref{lem:sphere}. Let $H:D^2\to X^1$ denote the map that realizes the homotopy between $p$ and $q$. We can assume $H$ is combinatorial as usual. Composing $H$ with the inclusion $X^1\subset X$ we can apply Lemma \ref{lem:key} to $H$ and obtain
$$
 T_p T_q^{-1} = T_{H(\partial D^2)} = \omega^{\tau^*H(D^2)}=I
$$
since the combinatorial map $H$ collapses all the cells of $D^2$.
}

As $\pi_2(X,X^1)$ is generated by the homotopy classes  of the elements $\gamma \tilde\Phi_j$  the image of $\partial$ is generated by $\partial(\gamma \tilde\Phi_j)$. We can  expand this element as a word in the generators $a_1,\cdots,a_k$ of the free group, say as a sequence $a_{i_1}^{\epsilon_1}\cdots a_{i_r}^{\epsilon_r}$ where $\epsilon_l=\pm 1$. Then applying Lemma \ref{lem:pathsinX1} to the homotopic paths $\partial(\gamma\tilde\Phi_j)$ and $a_{i_1}^{\epsilon_1}\cdots a_{i_r}^{\epsilon_r}$ and using Eq.~(\ref{eq:inv-path}) gives 
\begin{equation}\label{eq:looprelations}
\omega^{\Phi^*_j\tau(D^2)} =T_{\Phi_j(\partial D^2)} =  T_{\gamma\tilde\Phi_j(\partial D^2)} =\prod_{l=1}^{r} T_{a_{i_l}}^{\epsilon_l} .
\end{equation}
\vspace{5mm}

\Ex{\label{ex:looprel} {\rm
Let us express the loop relations Eq.~(\ref{eq:looprelations}) for the Mermin square, see Fig.~(\ref{fig:mermin-torus-projplane}). We will use the loops $a_i$ as defined in Example \ref{ex:TorusProj}. For notational simplification we write $T_i$ for $T_{a_i}$, and $\tau_j$ for $\tau(C_j)$ (also means $\Phi^*_j\tau(D^2)$).
We start with the torus case. With our choice of $a_i$ loops  the relations imposed by the characteristic maps can be written as
\begin{align*}
T_1 T_2 &= \omega^{\tau_1} \\
T_2^{-1} T_3 &= \omega^{\tau_2} \\
T_4 T_7^{-1} T_3^{-1} &= \omega^{\tau_3} \\ 
T_5^{-1}T_4^{-1} &= \omega^{\tau_4} \\
T_5 T_6^{-1} &= \omega^{\tau_5}\\
T_6T_7T_1^{-1} &= \omega^{\tau_6}. 
\end{align*}
If we multiply these equations in the given order then we end up with the equation
$$
[T_1,T_4] = \omega^{\sum_{i=1}^6 \tau_i}.
$$
We can verify this on Mermin square.  
In the projective plane case we get more information from the loop relations. They are given by
\begin{align*}
T_1 &= \omega^{\tau_1} \\
T_3 &= \omega^{\tau_2} \\
T_4T_7^{-1}T_3^{-1} &= \omega^{\tau_3} \\
T_4T_5^{-1} &= \omega^{\tau_4} \\
T_5 T_6^{-1} &= \omega^{\tau_5} \\
T_6T_7T_1&=\omega^{\tau_6}.
\end{align*}
Using these equations we obtain: $T_7 = \omega^{-\tau_2-\tau_3} T_4$, $T_5 = \omega^{-\tau_4}T_4$, $T_6=\omega^{-\tau_4-\tau_5}T_4$, $T_7= \omega^{\tau_4+\tau_5+\tau_6-\tau_1} T_4^{-1}$. Thus $T_i\sim_\omega T_4^{\pm 1}$ for $5\leq i\leq 7$. Comparing the two expressions for $T_7$ we obtain
\begin{equation}\label{eq:T4}
(T_4)^2 = \omega^{-\tau_1+\sum_{i=2}^6 \tau_i}.
\end{equation}
In particular we see that $T^0:\pi(X^1)\to U(n)$ has image given by an abelian group since all the $T_i$ commute with each other.
Now let us consider different cases for $d$:
\begin{itemize}
\item $d=2$: When $\omega=-1$ Eq.~(\ref{eq:T4}) becomes $T_4^2 = \pm 1$.  
First assume that $T_4=1$, in other words, $\sum \tau_i =0$. This means that $[\tau]=0$ in $H^2(X,\ZZ_2)$ and thus we have a classical realization. Note that the projective representation lifts to a linear representation in this case. Since each $T_i$ has order $2$ the image of $T^0$ is an elementary abelian $2$-group.   We can simply send 
$$
\rho[a_i] \mapsto T_i.
$$
This defines a group homomorphism $\tilde T:\pi_1(X)\to U(n)$.
For the other case where $T_4=-1$. We have $[\tau]\neq 0$ and thus no classical realization exists. This is the case for the Mermin star:
$$
(T_4)^2 = (X_1)(X_1X_2)(Z_2) = -1 
$$
where we used the usual operator realization.

\item $d$ odd: Eq.~(\ref{eq:T4}) has an interesting consequence. We can invert $2$ to obtain
$$
T_4 = \omega^{(-\tau_1+\sum_{i=2}^6 \tau_i)/2}
$$
and therefore  $T_i\sim_\omega I$ for all $i$. Therefore for any operator realization $T$ the loop operators turns out to be a scalar. To solve the classical realization problem we can use the assignment: $e_i \mapsto T_{a_i}$, which is a scalar, and the rest of the edges $e\notin \tT$ maps to $1$. This assignment satisfies the required relations and is a valid classical realization. From the representation point of view, we have that $T:\pi_1(X)\to U(n)/\Span{\omega}$ is the trivial representation for any operator realization $T$. Therefore a lift always exists.
\end{itemize}
}}
\vspace{5mm}

\Rem{{\rm
This illustrates the benefit of rewriting the operator relations as loop relations. As we also observed there is a strong connection between linearization of the projective representation and the existence of classical realization. Next we make this connection precise. 
}}
\vspace{5mm}

\Thm{\label{thm:lift-class}
The data $(L,\mM,\tau)$ together with an operator realization $T:L\to U(n)$ induces a projective representation $T:\pi_1(X)\to U(n)/\Span{\omega}$ such that the following are equivalent:
\begin{itemize}
\item There exists a classical realization of $(L,\mM,\tau)$.
\item $T$ can be lifted to a linear representation.
\end{itemize}
}
\Proof{
Suppose that $T$ lifts to a linear representation $\tilde T:\pi_1(X)\to U(n)$. By Lemma \ref{lem:linearization} for some $\alpha$ we have $T^{\alpha}(\im\,\partial)=I$. Using $\alpha$ let us define a classical realization $c:L\to \ZZ_d$ by setting $c(e_i)=-\alpha(a_i)$ for each edge $e_i$  corresponding to the generator $a_i$, and set $c(a)=0$ for the remaining labels in $L$. Recall that $a_i$ were originally defined form edges $e_i$ that lie outside of the maximal tree $\tT$.
 We claim that $c$ is a classical realization for $\tau$. We regard $c$ as an operator realization $C:L\to U(1)$. This way we can apply the results obtained for operator realizations. We need to check that $d c(C_j) = \tau(C_j)$. In terms of operator realizations this is equivalent to showing $C_{\Phi_j(\partial D^2)}=T_{\Phi_j(\partial D^2)}$ as a consequence of  Lemma \ref{lem:key} and the definition of $C$. Applying Eq.~(\ref{eq:looprelations}) to both $C$ and $T$ (taking $\gamma$ the trivial path) we can replace $\partial\tilde\Phi_j$ with the homotopic path $a_{i_1}^{\epsilon_1}\cdots a_{i_r}^{\epsilon_r}$. Now we calculate
\begin{equation}\label{eq:key}
\begin{aligned}
C^{-1}_{a_{i_1}^{\epsilon_1}\cdots a_{i_r}^{\epsilon_r}}\,T_{a_{i_1}^{\epsilon_1}\cdots a_{i_r}^{\epsilon_r}} =&  \prod_{l=1}^{r}  (\omega^{\alpha(a_{i_l})}T_{a_{i_l}})^{\epsilon_l} \\
=& T^{\alpha}(a_{i_1}^{\epsilon_1}\cdots a_{i_r}^{\epsilon_r})=I
\end{aligned}
\end{equation}
since $a_{i_1}^{\epsilon_1}\cdots a_{i_r}^{\epsilon_r}= \partial[\tilde\Phi_j]$ lies in the image of $\partial$.
 
Conversely, let $c:L\to \ZZ_d$ be a classical realization. Let $\alpha(a_i)=-c(e_i)$ and consider the operators $T^\alpha$.   Lemma \ref{lem:linearization} implies that to show that there exits a lift  it suffices to check that $T^\alpha(a_{i_1}^{\epsilon_1}\cdots a_{i_r}^{\epsilon_r})=I$ for any product $a_{i_1}^{\epsilon_1}\cdots a_{i_r}^{\epsilon_r}$ that lies in the image of $\partial$. Since $c$ is a classical realization   Lemma \ref{lem:key} implies that  
$
C_{a_{i_1}^{\epsilon_1}\cdots a_{i_r}^{\epsilon_r}} =T_{a_{i_1}^{\epsilon_1}\cdots a_{i_r}^{\epsilon_r}}
$.
Using Eq.~(\ref{eq:key}) we find that $T^\alpha(a_{i_1}^{\epsilon_1}\cdots a_{i_r}^{\epsilon_r})=I$.
  
}
\vspace{5mm} 
 
\Cor{\label{cor:generalization} If $(L,\mM)$ has a topological realization $X$ such that $\pi_1(X)$ is a finite group whose order is coprime to $d$ then any operator realizable $\tau$ is classically realizable.}
\Proof{
Let $T$ be an operator realization of $\tau$. By Theorem \ref{thm:lift-class} it suffices to show that there exists a lift $\tilde T: \pi_1(X) \to U(n)$. Let $G$ denote the quotient group $T(\pi_1(X))$. Similarly $G$ is a finite group and its order is coprime to $d$. To see that the lift $\tilde T$ exists it suffices to show that any extension
\begin{equation}
1\to \Span{\omega} \to \tilde G \to G \to 1
\end{equation}
splits. The extension splits as a consequence of the Schur-Zassenhaus theorem \cite[Theorem 7.41]{Rotman}. \change{This theorem says that when the order of the quotient group $G$ is coprime to the order of the kernel $\Span{\omega}$ then every extension splits.}
}

As we have observed in Example \ref{ex:looprel} that for projective plane realization of the Mermin square and $d$ is odd any operator relation has a classical solution. From the corollary we see that this is, in fact, a direct consequence of the fact that $\pi_1(\RR P^2)=\ZZ_2$ which implies $
H^2(\RR P^2 , \ZZ_d) = 0
$
when $d$ is odd.

\section{Quantum realizations and the  algorithmic approach} \label{sec:Quantum RealAlgorithmicApp}

\subsection{Quantum realization} We have defined the operator realization  without the requirement that the operators in each context must commute. Therefore  a quantum realization is an operator realization such that for each context $C\in \mM$ the operators $\set{T_a|\;a\in C}$ pairwise commute.  
We would like to understand the effect of the commutativity requirement on the topological realizations. 

\change{We will modify our definition of a topological realization to accommodate the extra requirement imposed by commutativity.}
For an operator realization we constructed a cell complex $\Xsing$ with a single vertex. Given the constraints the definition of this cell complex is unique. Then we defined a topological realization $X$ to be a cell complex with the same set of $1$-cells and $2$-cells but except possibly more than just a single vertex. The way the $2$-cells are glued in the new complex must be compatible with $\Xsing$ i.e. \change{comes with a map of cell complexes $
X\to \Xsing.
$}
Under this map the multiple vertices are identified to a single vertex.

When we pass to quantum realizations, \change{due to the commutativity condition,} the definition of $\Xsing$ is no longer unique since we can always permute the labels in a given context. Alternatively if we fix $\Xsing$ then we can   allow a more flexible notion of a topological realization that takes commutativity into account. \change{We choose the latter approach.}
\medskip

\Def{{\rm
\change{
Given a quantum realization and a fixed choice of $\Xsing$, a \textit{commutative topological realization} is a cell complex $X$ together with a map of chain complexes
$$
r_*:C_*(X) \to C_*(\Xsing)
$$
such that $r_1:C_1(X) \to C_1(\Xsing)$ and $r_2:C_2(X) \to C_2(\Xsing)$ are isomorphisms. 
}}
}

As it is reflected in this definition the commutation requirement ``abelianizes" the geometric requirement of having a map of cell complexes to  just having a map of chain complexes.  \change{This relaxation amounts for the freedom of permuting the labeling within contexts when we realize it topologically. To compare the two versions, for a fixed choice of $\Xsing$ a topological realization asks for a map of cell complexes $X\to \Xsing$, whereas a commutative topological realization only requires a map of chain complexes $C_*(X)\to C_*(\Xsing)$. }
 
\subsection{Orientation reversing}
A typical example of a commutative topological realization that is not allowed in the operator realization case     can be obtained by reversing the orientation of each cell. For a cell complex $X$ let us denote by $\bar X$ the cell complex obtained from $X$ by reversing the orientation of its edges, and reversing the orientation of its $2$-cells. If $X\to \Xsing$ is a cellular map, contracting all vertices to a single vertex, there may not be a cellular map $\bar X\to \Xsing$. But there is a map of chain complexes $C_*(\bar X) \to C_*(\Xsing)$, send a generator corresponding to an edge or a face to its negative. Therefore $X\mapsto \bar X$ is an operation that is available only for the quantum realizations. 
\vspace{5mm}

\Ex{{\rm
Let us consider the Mermin square example. In Fig.~(\ref{fig:orientation-reverse}) we see two choices of orientations for the cells. Let $\Xsing$ denote the complex obtained from $X$ by identifying all the vertices. We see that $\bar X$ does not admit a cellular map to $\Xsing$. The reason is that, for example, the top cell is glued in the wrong order: $A C D^{-1}$ versus $AD^{-1}C$. But when we pass to chain complexes the order does not matter, and a chain map exists. Next we illustrate an application of the    orientation reversing operation to deduce that the Mermin square arrangement is non-magic for $d$ odd. Suppose that there is a quantum realization $T$ for a given $\tau$. Note that on a $2$-cell $\Phi_j:D^2\to X$ and its orientation reversed version $\bar \Phi_j: D^2\to \bar X$ the evaluation of $\tau$ gives the same result since the operators placed on the edges of the boundary pairwise commute. Therefore evaluation of $\tau$ on the whole surface is the same in both cases. Let us write $\tau(X)$ for this common value. At the boundary we get two different equations
$$
A B A^{-1} B^{-1} = \omega^{\tau(X)},\;\;\; BAB^{-1} A^{-1}  =\omega^{\tau(X)}.
$$   
Combining them we obtain 
$$
\omega^{2\tau(X)} =1.
$$
But when $d$ is odd, this equation implies that $\tau(X)=0$. Equivalently, the arrangement is non-magic.

\begin{figure}[h!] \center
  \includegraphics[scale=0.7]{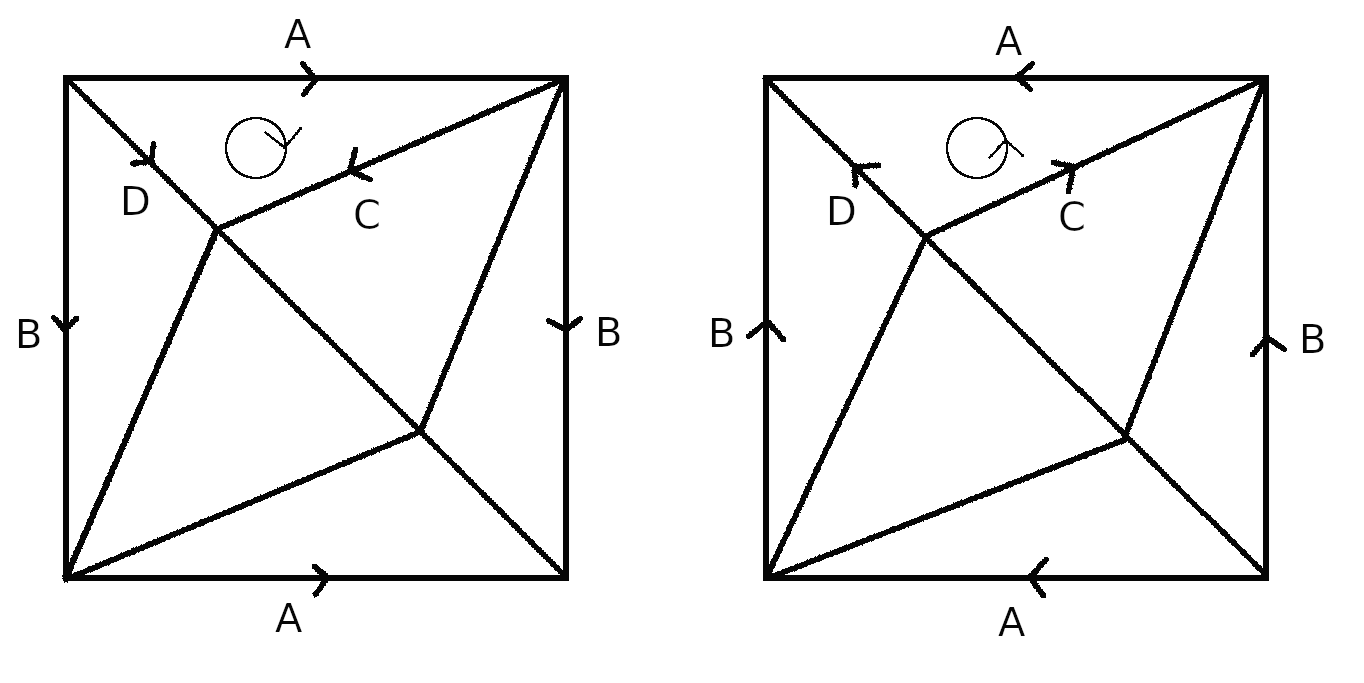}
  \caption{Reversing the orientation of the cells.}
  \label{fig:orientation-reverse}
\end{figure}  

}}

\subsection{Raising constraints to a power} Another special feature of quantum realizations is that given a constraint 
$$
\prod_{i=1}^k T_{a_i}^{\epsilon(a_i)} = \omega^{\tau(C)}
$$
and a natural number $m$ we can take the $m$-th power of both sides. Using commutativity we  obtain a new constraint
$$
\prod_{i=1}^k T_{a_i}^{m\epsilon(a_i)} = \omega^{m\tau(C)}.
$$
This is a very useful property which allows one to concentrate on a prime divisor of $d$. \change{Thus if $\set{T_a}$ is a quantum realization for $(L,\mM,\epsilon,\tau)$ then also $\set{T_a^m}$ is a quantum realization for the same arrangement.}

\change{We are interested in quantum realizations with $T_a^d=I$ for a fixed $d$.}
To emphasize dependence on $d$ let us denote a quantum realization by $(T,d)$. Let $p_i$ denote the prime divisors of $d$ so that we can write 
$$
d =\prod_{i=1}^t p_{i}^{\alpha_i}.
$$
For a fixed $j$ let us write $d_j$ for the product of $p_i^{\alpha_i}$ where $i\neq j$. \change{So that we have $d_j= d/p_j^{\alpha_j}$.} We can raise each operator $T_a$ to the $d_j$-th power $T_a^{d_j}$. We will write $(T^{d_j},p_j^{\alpha_j})$ for the resulting quantum realization. Our construction was in such a way that the new operators have order given by a prime power. \change{This follows from 
$$
(T_a^{d_j})^{p_j^{\alpha_j}} = (T_a)^{d_jp_j^{\alpha_j} }=T_a^d=I.
$$} Similarly  $\tau$ can be broken into prime power pieces. For a context $C$ we define $\tau_j(C)$ to be the product $d_j \tau(C)$. We write $\tau_j$ for the resulting assignment $\mM\to \ZZ_{p_j^{\alpha^j}}$.

Our goal is to ``glue" a given set of classical realizations for each ``prime" piece \change{$\tau_j$} to construct a classical realization for \change{$\tau$}.
\vspace{5mm}

\Pro{\label{pro:prime-reduction}
Given $(L,\mM,\tau)$ and a quantum realization $T$ the following are equivalent
\begin{itemize}
\item \change{$\tau$}  is classically realizable;
\item \change{$\tau_j$} is classically realizable for each prime divisor $p_j$ of $d$.
\end{itemize}
}
\Proof{One direction is immediate. Namely given a classical realization $c:\mM \to \ZZ_d$ for \change{$\tau$} we can simply set $c_j(C)=d_j c(C)$ to obtain a classical realization for \change{$\tau_j$} as a consequence of pairwise commutativity of operator constrains in each context.

Conversely, given classical realizations $c_j$ we would like to construct a global   $c$ that works for $d$. \change{Recall that we have $d =\prod_{i=1}^t p_{i}^{\alpha_i}$. The abelian group $\ZZ_d$ factors into a product $\ZZ_{p_1^{\alpha_1}}\times \cdots \times \ZZ_{p_t^{\alpha_t}}$. (For example $\ZZ_6=\ZZ_2\times \ZZ_3$.) Regard $d_j$, after reducing mod $p_j^{\alpha_j}$, as an element of $\ZZ_{p_j^{\alpha_j}}$. Therefore $\theta=(d_1,\cdots,d_t)$ specifies an element in $\ZZ_d$. We want to construct a multiplicative inverse to $\theta$. This can be done component-wise. Let $w_j$ be the multiplicative inverse of $d_j$ in the group $\ZZ_{p_j^{\alpha_j}}$. Then the element $w=(w_1,\cdots,w_t)$ is the inverse of $\theta$ i.e. $\theta w=1$ in $\ZZ_d$. This equation can also be written as} 
$$
\sum_{i=1}^t d_i w_i =1 \mod d. 
$$
Therefore we can write 
$$
T_a= \prod_{i=1}^t (T_a^{d_i})^{w_i},\;\;\;\; \tau = \sum_{i=1}^t  w_i\tau_i.
$$
Hence the classical realization defined by
$$
c= \sum_{i=1}^t w_i c_i
$$
will be a classical realization for \change{$\tau$}.
}

As a consequence of this observation for quantum realizations we can always restrict to the prime power case. This restriction also simplifies any algorithmic approach that checks whether a given arrangement is magic or not. Next we will explain the interplay between the topological and algorithmic approaches. We start with the relatively simpler scenario of   one-relator groups.

\subsection{One-relator groups} A convenient way of describing a group is to write down a presentation. \change{
A presentation consists of a set of generators and a set of relations.}  \change{If there is a presentation consisting of  only a single relation then the group is called a one-relator group.} For example, the fundamental sequence in Eq.~(\ref{eq:fundseq}) is a presentation of the fundamental group $\pi_1(X)$. This group is not necessarily a one-relator group. When we restrict our setting to Arkhipov's where only certain types of arrangements are allowed the fundamental group becomes a surface group, in particular a one-relator group. 

Let us suppose each observable belongs to exactly two contexts for the moment. Then the intersection \change{graph} $\gG$ can be embedded into a closed surface $\Sigma_g$ of minimal genus. We can use this surface as a topological realization in our framework. Its fundamental group is given by
$$
\pi_1(\Sigma_g) = \Span{a_1,b_1,\cdots,a_g,b_g|\; [a_1,b_1]\cdots [a_g,b_g]=1}
$$
and we have a single relation given by the vanishing of the product of the commutators. Moreover we can apply Lemma \ref{lem:key} to this situation. The surface $\Sigma_g$ can be obtained from a disk whose edges are labeled by $a_i,b_i,a_i^{-1},b_i^{-1}$ where $1\leq i\leq g$. We can think of this as a (combinatorial) map
$$
 D^2 \to \Sigma_g
$$
identifying the boundary with the sequence of loops in the surface described by the partitioning of the boundary. For a set of operator constraints Lemma \ref{lem:key} applies to give
$$
[T_{a_1},T_{b_1}] \cdots [T_{a_g},T_{b_g}] = \omega^{\tau(\Sigma_g)}
$$
that is evaluation of $\tau$ on the whole surface can be computed from the product of the operators on the boundary, which is given by the product of the commutators. Note that this is the relation defining the fundamental group of the surface realizing the operator constraints. Alternatively, we can read this equality in terms of the product of the constraints
\begin{equation}\label{eq:com-prod}
[T_{a_1},T_{b_1}] \cdots [T_{a_g},T_{b_g}] = \prod_{C\in \mM} \left( \prod_{a\in C} T_a \right).
\end{equation}
Now the restriction that each observable belongs to two contexts (or more generally even number of contexts) implies that the existence of a classical realization forces the right-hand side of Eq.~(\ref{eq:com-prod}) to be equal to $1$. \change{This is an AvN type of argument \cite{abramsky2015contextuality}.}

\subsection{Solution group} It is possible to assign an abstract group for a set of operator constraints \cite{Slo,KSpaper}. We will consider the quantum case by imposing the commutativity constraint in each context. Recall that $L$ denotes the label set, i.e. the union of the contexts in  $\mM$, also corresponds to the set of edges of a topological realization, \change{and $\tau$ is a function $\mM\to\ZZ_d$.} The solution group $G$  is generated by $g_a$ where $a\in L$, and $J$ satisfying
\begin{itemize}
\item $J^d=(g_a)^d =1$ for all $a\in L$;
\item \change{$[J,g_a]=1$} for all $a\in L$;
\item $[g_a,g_b]=1$ for all $a,b\in C$ and $C\in \mM$;\footnote{For operator realizations we remove this condition.}
\item For each $C\in \mM$ there is a relation of the form
$$
\prod_{a\in C} g_a = J^{\tau(C)}.
$$
\end{itemize}
We denote the quotient group $G/\Span{J}$ by $\bar G$. We can think of this as a central extension of groups
$$
1\to \Span{J} \to  G \to \bar G \to 1.
$$ 
Note that $J$ replaces the role of $\omega I$ in the unitary group.  

Going back to the restricted setting of Arkhipov the algorithmic approach \change{as explained in \cite{KSpaper}} decides about existence of classical realizations by checking the right-hand side of Eq.~(\ref{eq:com-prod}) in the solution group. That is existence of a classical realization is equivalent to the condition that the product of the constrains satisfy
\begin{equation}\label{eq:restricted-eq}
\prod_{C\in\mM} \left( \prod_{a\in C} g_a \right)=1\;\text{ in }G,
\end{equation}
where it is also assumed that $d=2$. The equivalence of the two conditions is proved in \cite[Proposition 10 \& 11]{Arkhipov}.

We would like to generalize this result  by making a connection to Theorem \ref{thm:lift-class}. The projective representation defined there can be modified by replacing the unitary group $U(n)$ by the solution group. Given a (combinatorial) path $p$ specified by a sequence of edges $a_i^{\epsilon(a_i)}$ in $X$ we define the group element
$$
g_p = \prod_{i=1}^k g_{a_i}^{\epsilon(a_i)}.
$$
Then the assignment $[p]\mapsto g_p$ gives a group homomorphism
$$
\theta: \pi_1(X) \to \bar G.
$$
Then Theorem \ref{thm:lift-class} takes the following form.
\vspace{5mm}

\Cor{\label{cor:lift}
The following conditions are equivalent:
\begin{itemize}
\item There exists a classical realization for the set of constraints;
\item $\theta:\pi_1(X)\to \bar G$ lifts to a homomorphism $\tilde \theta: \pi_1(X)\to G$.
\end{itemize}
}
 
This result is the generalization of the equivalence between the existence of a  classical realization and the  product of all the constraints being equal to $1$, that holds in the restricted setting of Arkhipov. 
To illustrate this we consider the one-relator case of surface groups. Eq.~(\ref{eq:com-prod}), obtained from Lemma \ref{lem:key}, also holds in $G$, that is we have
\begin{equation}\label{eq:com-prod-G}
 \prod_{i=1}^g\, [g_{a_i},g_{b_i}]=\prod_{C\in\mM} \left( \prod_{a\in C} g_a \right) 
\end{equation}
since its derivation does not depend on the unitary group.  Note that the lift $\tilde \theta$ exists if and only if the right-hand side of Eq.~(\ref{eq:com-prod-G}) is equal to $1$. On the other hand, a classical realization exists if and only if the right-hand side gives us $1$. Thus this illustrates Corollary \ref{cor:lift} in the restricted setting of Arkhipov.

\subsection{General case} So far we have looked at the relationship between the topological approach and the algorithmic approach in the restricted types of contexts and the case where $d=2$. For arbitrary $d$ and arbitrary collection of contexts the algorithmic approach breaks down. That is checking the value of the product of all the constraints in the solution group does not indicate anything about classical realizability.

Our approach solves this problem. We can interpret Corollary \ref{cor:lift} as follows. For a fixed choice of topological realization $X$ the fundamental group $\pi_1(X)$ has a specific presentation
$$
\pi_1(X) = \Span{a_1,\cdots,a_n|\; r_1,\cdots,r_m }
$$
with generators $a_i$ and relations $r_j$. In the general case the fundamental group is not necessarily a one-relator group i.e. $m\geq 1$. Corollary \ref{cor:lift} says that we need to check whether $\tilde \theta$ lifts to $G$, that is equivalently we need to check whether the relations $r_i$ also hold in $G$. If we write $r_i$ as an equation $w_i=1$ for some word $w_i$ then one needs to verify in the group $G$ the following set of equations:
\begin{equation}\label{eq:relations-G}
\tilde \theta(w_i) =1 \;\text{ where } 1\leq i\leq m.
\end{equation}
This can be checked, for example, using the Knuth-Bendix algorithm \cite[Chapter 12]{CompGroup}. There is an extra freedom here, that is the choice of $X$. A different topological realization will change the relations $r_i$ to some other relations. Among these different choices some of them would be more favorable in verifying Eq.~(\ref{eq:relations-G}) algorithmically. 

\section{Conclusion} 

The goal of this work is to extend Arkhipov's result on magic arrangements. In the restricted scenario where each observable belongs to exactly two contexts Arkhipov studied contextuality using the  intersection graph. For arbitrary arrangements this graph theoretical tool is not available. We have overcome this difficulty by introducing the notion of a topological realization. In this formulation we showed that an arrangement is magic, in other words contextual, only if the fundamental group of the topological realization is non-trivial. We refined this approach further by defining a projective representation of the fundamental group. A classical realization for the given set of constraints can be regarded as a linearization of the representation. When the fundamental group is finite divisibility of its order by a prime divisor of $d$ is required in order to have a magic arrangement.  
Moreover, quantum realizations can be studied one prime at a time, that is we loose no generality by restricting $d$ to be a prime power.
 All these observations can be combined with the algorithmic approach that can be utilized by the Knuth--Bendix algorithm for instance. Dialing through different topological realizations produce different sets of relations, generalizing one-relation surface groups, whose verification is expected to vary in efficiency. Thus both approaches can be combined to study magic arrangements.

\paragraph{Acknowledgments.} This work is supported by NSERC (CO,RR), Cifar (RR), the Stewart Blusson Quantum Matter
Institute (CO).


\onecolumn\newpage
\appendix
 
\section{Appendix: Topology of $2$-dimensional complexes \label{sec:app}} 
 
In this section we describe some properties of cell complexes. In particular, we describe combinatorial complexes and combinatorial maps, which are the main technical tools to study the topology of $2$-dimensional cell complexes. 
\change{{\bf Notation:} Following \cite{Hatcher} we write $A\subset B$ or $A\supset B$ for set-theoretic containment, not necessarily proper. As a map an inclusion (monomorphism) is denoted by $\hookrightarrow$ and a surjection (epimorphism) is denoted by $\twoheadrightarrow$.}
 
\subsection{Cell complexes} Let $D^n$ denote the unit disk in $\RR^n$, and $S^n$ denote its boundary $\partial D^n$.

A cell complex is constructed in the following way \cite{Hatcher}:
\begin{enumerate}
\item Start with a set $X^0$, called the set of $0$-cells.
\item Construct the $n$-skeleton $X^n$ by attaching $n$-cells to $X^{n-1}$
via the attaching maps $\phi_j:\partial D^n_j\to X^{n-1}$.
\end{enumerate} 

Each cell has a characteristic map denoted by $\Phi_j :D^n_j \to X$. \change{A typical example is the following cell structure of a torus $T$. Let $T^0$ consist of a single point $\set{\ast}$. Glue two $1$-cells  using the attaching maps  $\partial D^1 \to T^0$ by sending both boundary points to $\ast$. The resulting object is the $1$-skeleton $T^1$ and it is a bouquet of two circles, call these two loops $\alpha$ and $\beta$. Finally glue a $2$-cell using the attaching map $\partial D^2\to T^1$ by sending the boundary to the loop $\alpha \cdot \beta\ \cdot\alpha^{-1}\cdot \beta^{-1}$.}

A map $f:X\to Y$ between cell complexes is called cellular if $f(X^n)\subset Y^n$. It is well known that any map is homotopic to a cellular map. For our purposes we need maps that have nicer properties than cellular maps in the sense that they are more combinatorial.

\subsection{Combinatorial maps} We will describe combinatorial maps only for $2$-complexes \cite[Chapter II, \S 1.2]{TwoDim}. \change{These types of cell complexes are more rigid in structure and can be manipulated much easier. We need the notion of a product cell.} A $2$-cell is a product cell if it is described by a map $\phi:D^1\times D^1\to X$ such that $\phi(D^1\times \set{1})$ and $\phi(D^1\times \set{-1})$ are $1$-cells of $X$. \change{Product cells can be collapsed in a special way. Let $f:X\to Y$ be a cellular map. This map  collapses the product cell $\phi$ if $f$ restricts to the same function on $\phi(D^1\times \set{t})$ for all $t\in D^1$. Thus the $2$-cell is collapsed onto the image of the $1$-cells $\phi(D^1\times \set{\pm 1})$.}

We say a cellular map $f:X\to Y$ is combinatorial if
\begin{itemize}
\item Every $1$-cell of $X$  is either mapped homeomorphically onto a $1$-cell of $Y$ or collapsed to a $0$-cell

\item Every $2$-cell of $X$ is either mapped homeomorphically onto a $2$-cell of $Y$ or it is a product cell that is collapsed to a $1$-cell or a $0$-cell. 

\end{itemize}

A $2$-complex is called combinatorial if the attaching map of every \change{$2$-cell} is combinatorial. \change{In this type of cell complexes the behavior of the attaching maps is simpler. But still the homotopy type captures all possible $2$-dimensional cell complexes. Topological realizations introduced in the text are examples of combinatorial complexes. The rigidity of such complexes will be immensely useful to understand the relationship between path operators and the cocycle $\tau$ determining the product constraints. Up to homotopy cellular maps are always combinatorial as a consequence of the following technical Lemma.}
 \vspace{5mm}
 
\Lem{\label{lem:uptohomotopy} \cite[Chapter II Theorem 1.5, Theorem 1.8]{TwoDim} Let $X$ be a combinatorial $2$-complex.
\begin{itemize}
\item  For a given map $f:S^1\to X^1$ there is a cell structure on $S^1$ such that $f$  is homotopic to a combinatorial map. 

\item  Let $g:D^2\to X$ be a map such that the restriction to the boundary is combinatorial. Then there is a cell structure on $D^2$, extending the one on its boundary, such that $g$ is homotopic (relative to the boundary) to a combinatorial map. 
\end{itemize}
}

\end{document}